\documentclass[amsmath,aps,pra,preprint,showpacs]{revtex4-1}
\usepackage{graphics}

\begin{document} 

\title{Cavity-enhanced channeling of emission from an atom into a nanofiber}

\author{Fam Le Kien}
\altaffiliation{Also at Institute of Physics, Vietnamese Academy of Science and Technology, Hanoi, Vietnam.}

\author{K. Hakuta} 

\affiliation{Department of Applied Physics and Chemistry, 
University of Electro-Communications, Chofu, Tokyo 182-8585, Japan}

\date{\today}

\begin{abstract}
We study spontaneous emission of an atom near a nanofiber with two fiber-Bragg-grating (FBG) mirrors.
We show that the coupling between the atom and the guided modes of the nanofiber can be significantly enhanced by the FBG cavity even when the cavity finesse is moderate. We find that, when the fiber radius is 200 nm and the cavity finesse is about 30, up to 94\% of spontaneous emission from the atom can be channeled into the guided modes in the overdamped-cavity regime. We show numerically and analytically that vacuum Rabi oscillations and strong coupling can occur in the FBG cavity even when the cavity finesse is moderate (about 30) and the cavity length is large (on the order of 10 cm to 1 m), unlike the case of planar and curved Fabry-P\'{e}rot cavities.
\end{abstract}

\maketitle

\section{Introduction}

Coupling of light to subwavelength structures and its control pose one of the greatest challenges of recent research \cite{Maier,Wallraff,Lukin,cesium decay,magnetic,Kali,Nayak}. Strong coupling in a superconducting circuit at microwave frequencies has been observed \cite{Wallraff}. Chang \textit{et al.} have proposed a technique that enables strong coherent coupling between individual emitters and guided plasmon excitations in conducting nanostructures \cite{Lukin}. In the case of \textit{dielectric} waveguides, it has been shown that a significant fraction (up to 28\%) of emission from a single atom can be channeled into a nanofiber \cite{cesium decay,magnetic,Kali,Nayak}.
Radiative decay of an atom in the vicinity of a nanofiber has been studied in the context of a two-level atom \cite{Jhe,Tromborg,Klimov} as well as a realistic multilevel cesium atom with a hyperfine structure of energy levels \cite{cesium decay,magnetic}. The parameters for the decay of populations and cross-level coherences of an atom near a nanofiber have been calculated \cite{cesium decay,magnetic}. The cooperation of distant atoms via a nanofiber has been discussed \cite{two atoms,multiatom}. It has been shown that, due to guided modes, a substantial cooperation can survive large interatomic distances \cite{two atoms}, and a linear array of distant atoms can significantly enhance the rate of spontaneous emission and the efficiency of channeling of emission into the nanofiber \cite{multiatom}.

Optical cavities are often employed to increase the interaction between atoms and photons 
\cite{Berman,Thompson,Mabuchi,Kimble group,Rempe,Shimizu,McKeever,Maunz,Sauer,Aoki,Zhu2007,Muller,Lukin1998,Xiao2000,Xiao2008,Zhu2008,Black}. Various cavity quantum electrodynamic effects have been studied \cite{Berman}. 
There have been spectacular recent successes brought by the merging of optical cavity
systems with ultracold neutral atoms \cite{Mabuchi,Kimble group,Rempe,Shimizu,McKeever,Maunz,Sauer,Aoki}
as well as with electromagnetically induced transparency physics 
\cite{Lukin1998,Muller,Xiao2000,Zhu2007,Xiao2008,Zhu2008,Black}. It is natural to expect that
the use of a cavity can substantially enhance the channeling of emission from an atom into a nanostructure. It is desirable to combine the cavity technique with the nanofiber technique to obtain a hybrid system, where the interaction is enhanced by the transverse confinement of the field in the fiber cross-section plane as well as the longitudinal confinement of the field between the mirrors. 
Such a system has been studied recently in the context of intracavity electromagnetically induced transparency \cite{fibercavity}. It has been shown that the presence of a fiber-Bragg-grating (FBG) cavity with a large length (on the order of 10 cm) and a moderate finesse (about 30) can significantly enhance the group delay of the guided probe field \cite{fibercavity}.

In this paper, we study spontaneous emission of an atom near a nanofiber with two FBG mirrors. We find that the coupling between the atom and the guided modes can be significantly enhanced by the FBG cavity even when the cavity finesse is moderate. We show numerically and analytically that vacuum Rabi oscillations and strong coupling can occur in the FBG cavity even when the cavity finesse is moderate (about 30) and the cavity length is large (on the order of 10 cm to 1 m). 

Before we proceed, we note that there has been a large body of work involving fiber Bragg gratings over the past two decades \cite{Othenos,Kashyap,Canning,Wan,Gupta,Chow}. With careful control of the grating writing process and appropriate choice of glass material, a FBG resonator can have a finesse of well over 1000 and a linewidth of a few MHz \cite{Gupta}. 

The paper is organized as follows. In Sec.\ \ref{sec:model} we describe the model of a nanofiber with
two FBG mirrors. In Sec.\ \ref{sec:basic} we derive a basic equation for spontaneous emission of an atom in the model. In Sec.\ \ref{sec:markov} we study spontaneous emission of the atom in the overdamped-cavity regime. In Sec.\ \ref{sec:delay} we derive a delay-differential equation for spontaneous emission and study it numerically. In Sec.\ \ref{sec:single mode} we approximate the delay-differential equation under the single-mode cavity condition and analyze the atomic decay in various cases. Our conclusions are given in Sec.~\ref{sec:summary}.

\section{MODEL}
\label{sec:model}

\begin{figure}[tbh]
\begin{center}
 \includegraphics{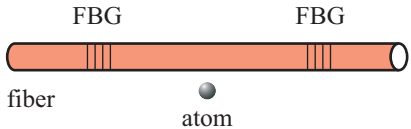}
 \end{center}
\caption{(Color online) Spontaneous emission of an atom in the vicinity of a nanofiber with two 
fiber-Bragg-grating mirrors.}
\label{fig1}
\end{figure}

We consider spontaneous emission of a two-level atom in the vicinity of a nanofiber with two FBG mirrors (see Fig. \ref{fig1}). The field in the guided modes of the nanofiber is reflected back and forth between the FBG mirrors. Such a system is a nanofiber-based cavity. The nanofiber has a cylindrical silica core of radius $a$ and of refractive index $n_1=1.45$ and an infinite vacuum clad of refractive index $n_2=1$. In view of the very low losses of silica in the wavelength range of interest, we neglect material absorption. We also neglect the effects of the surface-induced potential,
the surface roughness, and the phonon heating on the atom. We use the cylindrical coordinates $(r,\varphi,z)$, with $z$ being the axis of the fiber.

%%%%%%%%%%%%%%%%%%%%%%% Figure 1

In the presence of the fiber, the electromagnetic field can be decomposed into guided and radiation modes \cite{fiber books}. In order to describe the field in a quantum mechanical treatment, we follow the continuum field quantization procedures presented in \cite{Loudon}. First, we temporally neglect the presence of the FBG mirrors. 
Regarding the guided modes, we assume that the single-mode condition \cite{fiber books} 
is satisfied for a finite bandwidth around the atomic transition frequency $\omega_0$. 
We label each fundamental guided mode HE$_{11}$ with a frequency $\omega$ in this bandwidth by an index $\mu=(\omega,f,l)$, 
where $f=+,-$ denotes the forward or backward propagation direction
and $l=+,-$ denotes the counterclockwise or clockwise rotation of polarization. In the interaction picture, the quantum expression for the electric positive-frequency component 
$\mathbf{E}^{(+)}_{\mathrm{gyd}}$ of the field in the cavity-free guided modes is \cite{cesium decay}
\begin{equation}
\mathbf{E}^{(+)}_{\mathrm{gyd}}=i\sum_{\mu}\sqrt{\frac{\hbar\omega\beta'}{4\pi\epsilon_0}}
\;a_{\mu}\mathbf{e}^{(\mu)}e^{-i(\omega t-f\beta z-l\varphi)}.
\label{3n}
\end{equation}
Here $\mathbf{e}^{(\mu)}=\mathbf{e}^{(\mu)}(r,\varphi)$ is the profile function of the guided mode $\mu$ in the classical problem, $a_{\mu}$ is the corresponding photon annihilation operator, 
$\sum_{\mu}=\sum_{fl}\int_0^{\infty}d\omega$ is the summation over the guided modes,
$\beta$ is the longitudinal propagation constant, and $\beta'$ is the derivative of $\beta$
with respect to $\omega$. The constant $\beta$ is determined by the
fiber eigenvalue equation \cite{fiber books}. The operators $a_{\mu}$ and $a_{\mu}^\dagger$ satisfy the continuous-mode bosonic commutation rules $[a_{\mu},a_{\mu'}^\dagger]=\delta(\omega-\omega')\delta_{ff'}\delta_{ll'}$. The explicit expression for the mode function $\mathbf{e}^{(\mu)}$ is given
in Appendix \ref{sec:guided} (see also Refs. \cite{cesium decay,fiber books}). 
 
Regarding the radiation modes, the longitudinal propagation constant $\beta$ for each frequency $\omega$ can vary continuously, from $-k$ to $k$, with $k=\omega/c$ being the wave number. 
We label each radiation mode by an index $\nu=(\omega,\beta,m,l)$, where 
$m=0,\pm1,\pm2,\dots$ is the mode order and $l=+,-$ is the mode polarization. In the interaction picture, the quantum expression for the electric positive-frequency component 
$\mathbf{E}^{(+)}_{\mathrm{rad}}$ of the field in the radiation modes is \cite{cesium decay}
\begin{equation}
\mathbf{E}^{(+)}_{\mathrm{rad}}=i\sum_{\nu}
\sqrt{\frac{\hbar\omega}{4\pi\epsilon_0}}\;a_{\nu}\mathbf{e}^{(\nu)}e^{-i(\omega t-\beta z-m\varphi)}.
\label{4n}
\end{equation}
Here $\mathbf{e}^{(\nu)}=\mathbf{e}^{(\nu)}(r,\varphi)$ is the profile function of the radiation mode $\nu$ in the classical problem, $a_{\nu}$ is the corresponding photon annihilation 
operator, and $\sum_{\nu}=\sum_{ml}\int_0^{\infty}d\omega\int_{-k}^{k}d\beta$ is the summation over the radiation modes. The operators $a_{\nu}$ and $a_{\nu}^\dagger$ satisfy the continuous-mode bosonic commutation rules $[a_{\nu},a_{\nu'}^\dagger]=\delta(\omega-\omega')\delta(\beta-\beta')
\delta_{mm'}\delta_{ll'}$. The explicit expression for the mode function $\mathbf{e}^{(\nu)}$ is given
in Appendix \ref{sec:radiation} (see also Refs. \cite{cesium decay,fiber books}).

Next, we take into account the effect of the FBG mirrors on the mode functions. 
We assume that the two FBG mirrors
are identical, having the same complex reflection and transmission coefficients $R$ and $T$, respectively, for the guided modes in a broad bandwidth around the atomic transition frequency $\omega_0$. In general, we have $|R|^2+|T|^2\leq 1$, where the equality (inequality) occurs for lossless (lossy) gratings. Without loss of essential physics, we assume that the gratings are lossless, that is, $|R|^2+|T|^2=1$. Let the mirrors be separated by a distance $L$, from 
the point $z=-L/2$ to the point $z=L/2$. The mode functions of the guided modes are modified by the
presence of the mirrors. The forms of the cavity-modified mode functions are obtained, as usual in the Fabry-P\'{e}rot theory, by summing the geometric series resulting from the multiple reflections by the mirrors
\cite{Martini,Bjork,Cook}. Inside the cavity, the mode functions of the cavity-modified guided modes are given by
\begin{eqnarray}\label{7}
\tilde{\mathbf{e}}^{(\omega,+,l)}&=&\mathbf{e}^{(\omega,+,l)}\frac{T}{1-R^2e^{2i\beta L}}
+\mathbf{e}^{(\omega,-,l)}\frac{TR e^{i\beta(L-2z)}}{1-R^2e^{2i\beta L}},
\nonumber\\
\tilde{\mathbf{e}}^{(\omega,-,l)}&=&\mathbf{e}^{(\omega,-,l)}\frac{T}{1-R^2e^{2i\beta L}}
+\mathbf{e}^{(\omega,+,l)}\frac{TR e^{i\beta (L+2z)}}{1-R^2e^{2i\beta L}},
\nonumber\\
\end{eqnarray}
and, hence, the electric positive-frequency component of the field in the cavity-modified guided modes is
\begin{equation}
\mathbf{E}^{(+)}_{\mathrm{cavgyd}}=i\sum_{\mu}\sqrt{\frac{\hbar\omega\beta'}{4\pi\epsilon_0}}
\;a_{\mu}\tilde{\mathbf{e}}^{(\mu)}e^{-i(\omega t-f\beta z-l\varphi)}.
\label{3na}
\end{equation}

We assume that the FBG mirrors do not reflect the radiation modes. This assumption is reasonable in the case where the distance $L$ between the FBG mirrors is large as compared to the fiber radius $a$
and to the wavelength $\lambda_0=2\pi/k_0$, with $k_0=\omega_0/c$ being the wave number of the atomic transition. With this assumption, the mode functions of the radiation modes are unchanged by the presence of the FBG mirrors. Inside the cavity, the electric positive-frequency component of the total field is given by
\begin{equation}
\mathbf{E}^{(+)}=\mathbf{E}^{(+)}_{\mathrm{cavgyd}}+\mathbf{E}^{(+)}_{\mathrm{rad}}.
\label{3nb}
\end{equation}

We emphasize that the FBG cavity described above confines only the guided modes, whose wave vectors are aligned along the fiber axis direction $z$. The radiation modes are not confined by the FBG cavity. 
In this sense, the physics of the FBG cavity is similar to that of one-dimensional cavities \cite{Cook,Feng}, and is different from that of planar Fabry-P\'{e}rot cavities \cite{Berman,Martini,Bjork,Dung}, where off-axis modes reduce the quantum electrodynamic (QED) effect of the cavity on spontaneous emission of the atom \cite{Bjork,Dung}. We also note that 
the guided field in the FBG cavity is confined not only in the axial direction between the mirrors but also in the fiber cross-section plane. In this sense, the physics of the FBG cavity is similar to that of curved Fabry-P\'{e}rot cavities, which are often used in experiments on cavity QED effects \cite{Berman,Thompson,Mabuchi,Kimble group,Rempe,Shimizu,McKeever,Maunz,Sauer}. An advantage of a FBG cavity based on a nanofiber is that the field in the guided modes can be confined to a small cross-section area whose size is comparable to the light wavelength \cite{nanofiber properties}. 
For example, for a nanofiber with radius of 200 nm, the effective mode area 
$A_{\mathrm{eff}}=(\int |\mathbf{e}^{(\mu)}|^2d\mathbf{r})^2/\int |\mathbf{e}^{(\mu)}|^4d\mathbf{r}$ 
of the fundamental guided modes with the wavelength $\lambda=852$ nm is found to be $A_{\mathrm{eff}}\cong 0.65$ $\mu\mathrm{m}^2$. The corresponding mode radius is found to be $r_{\mathrm{eff}}=\sqrt{A_{\mathrm{eff}}/\pi}\cong 454$ nm, which
is much smaller than the typical values of 15 to 30 $\mu$m for the waists of the cavity modes used in the experiments on cavity QED effects \cite{Berman,Thompson,Mabuchi,Kimble group,Rempe,Shimizu,McKeever,Maunz,Sauer}. Another advantage of the nanofiber-based cavity is that the cavity guided field can be transmitted over long distances for the communication purposes.

We now describe the interaction between the atom and the field.
Let $|a\rangle$ and $|b\rangle$ be the upper and lower states of the atom, respectively.
The operators $\sigma=|b\rangle\langle a|$ and 
$\sigma^\dagger=|a\rangle\langle b|$
describe the downward and upward transitions of the atom, respectively. 
In the dipole and rotating-wave approximations and in the interaction picture, the Hamiltonian for the atom--field interaction is 
\begin{eqnarray}\label{20}
H_{\mathrm{int}}&=&-i\hbar\sum_{\mu}\tilde{G}_{\mu}\sigma^{\dagger} a_{\mu}
e^{-i(\omega-\omega_0)t}
\nonumber\\&&\mbox{}
-i\hbar\sum_{\nu}G_{\nu}\sigma^{\dagger} a_{\nu}
e^{-i(\omega-\omega_0)t}
+\mbox{H.c.},
\end{eqnarray}
where the coefficients $\tilde{G}_{\mu}$ and $G_{\nu}$ 
characterize the coupling of the atom with
the cavity-modified guided modes $\mu=(\omega, f, l)$ and the radiation modes $\nu=(\omega,\beta, m, l)$, respectively. 
Their explicit expressions are
\begin{subequations} 
\begin{eqnarray}\label{21}
\tilde{G}_{\mu}&=&\sqrt{\frac{\omega\beta'}{4\pi\epsilon_0\hbar}}\;
\big[\mathbf{d}\cdot\tilde{\mathbf{e}}^{(\mu)}(r,\varphi,z)\big]e^{i(f\beta z+l\varphi)},\label{21a}\\
G_{\nu}&=&\sqrt{\frac{\omega}{4\pi\epsilon_0\hbar}}\;
\big[\mathbf{d}\cdot\mathbf{e}^{(\nu)}(r,\varphi)\big]e^{i(\beta z+m\varphi)}.\label{21b}
\end{eqnarray}
\end{subequations}
Here $\mathbf{d}=\langle a|\hat{\mathbf{d}}|b\rangle$ is the matrix element of the electric dipole moment of the atom, and
$r$, $\varphi$, and $z$ are the cylindrical coordinates of the position of the atom.

\section{BASIC EQUATION FOR SPONTANEOUS EMISSION}
\label{sec:basic}

We assume that the atom is initially excited and the field is initially in the vacuum state. The wave function of the combined atom--field system at an arbitrary time $t$ can be written as
\begin{equation}\label{22}
|\psi\rangle=C_a|a;0\rangle+\sum_{\mu}C_{b\mu}|b;1_{\mu}\rangle
+\sum_{\nu}C_{b\nu}|b;1_{\nu}\rangle.
\end{equation}
Here $C_a$ is the probability amplitude for the atom to remain in the upper state $|a\rangle$, and
$C_{b\mu}$ and $C_{b\nu}$ are the probability amplitudes for the atom to move to the lower state $|b\rangle$, emitting a photon into
a guided mode $\mu$ and a radiation mode $\nu$, respectively. 
In the interaction picture, the Schr\"{o}dinger equation $i\hbar |\dot{\psi}\rangle=H_{\mathrm{int}}|\psi\rangle$ yields the following equations for the probability amplitudes:
\begin{equation}\label{23a}
\dot{C}_a=-\sum_{\mu} \tilde{G}_{\mu}e^{-i(\omega-\omega_0)t}C_{b\mu}
-\sum_{\nu} G_{\nu}e^{-i(\omega-\omega_0)t}C_{b\nu}
\end{equation}
and
\begin{eqnarray}\label{23b}
\dot{C}_{b\mu}&=& \tilde{G}_{\mu}^*e^{i(\omega-\omega_0)t}C_a,
\nonumber\\
\dot{C}_{b\nu}&=& G_{\nu}^*e^{i(\omega-\omega_0)t}C_a.
\end{eqnarray}
We integrate Eqs. (\ref{23b}) and substitute the results into Eq. (\ref{23a}).
Then, we obtain 
\begin{eqnarray}\label{24}
\dot{C}_a(t)&=& -\sum_{\mu} |\tilde{G}_{\mu}|^2 \int_0^t e^{-i(\omega-\omega_0)\tau}C_a(t-\tau) d\tau
\nonumber\\&&\mbox{}
-\sum_{\nu} |G_{\nu}|^2 \int_0^t e^{-i(\omega-\omega_0)\tau}C_a(t-\tau) d\tau.\qquad
\end{eqnarray}

Since the radiation modes are not confined by the FBG cavity, the interaction between the atom and the field in the radiation modes is weak. In addition, the mode functions $\mathbf{e}^{(\nu)}$ of the radiation modes are smooth with respect to the mode frequencies. Therefore, we can apply the Born-Markov approximation to the contribution of the radiation modes, that is, to the terms associated with the second integral on the right side of Eq. \eqref{24}. In this approximation, we replace $C_a(t-\tau)$ by $C_a(t)$ and take it out from the integral. With the assumption that the observation time $t$ is much larger than the atomic oscillation period $2\pi/\omega_0$, we extend the upper integration limit $t$ to $+\infty$. Furthermore, we neglect the imaginary part of the result of the integration, which describes the contribution of the radiation modes to the Lamb shift of the atomic transition frequency. Then, we obtain
\begin{equation}\label{25}
\dot{C}_a(t)=\big[\dot{C}_a(t)\big]_{\mathrm{gyd}}-\frac{\gamma_{\mathrm{rad}}}{2}C_a(t),
\end{equation}
where the term
\begin{equation}\label{26}
\big[\dot{C}_a(t)\big]_{\mathrm{gyd}}=-\sum_{\mu} |\tilde{G}_{\mu}|^2 \int_0^t e^{-i(\omega-\omega_0)\tau} C_a(t-\tau) d\tau
\end{equation}
describes spontaneous emission into guided modes
and the coefficient 
\begin{equation}\label{27}
\gamma_{\mathrm{rad}}=2\pi\sum_{\nu} |G_{\nu}|^2\delta(\omega-\omega_0)
\end{equation}
is the rate of spontaneous emission into radiation modes. 

In terms of the mode functions $\mathbf{e}^{(\nu)}$ of the radiation modes, expression (\ref{27}) for $\gamma_{\mathrm{rad}}$ can be rewritten as \cite{cesium decay,two atoms}
\begin{equation}\label{27a}
\gamma_{\mathrm{rad}}=\frac{\omega_0}{2\epsilon_0\hbar}\sum_{ml}\int _{-k_0}^{k_0}d\beta\,
\big|\mathbf{d}\cdot\mathbf{e}^{(\omega_0\beta m l)}(r,\varphi)\big|^2.
\end{equation}
The rate $\gamma_{\mathrm{rad}}$ of spontaneous emission into radiation modes has been calculated and studied in Refs. \cite{cesium decay,magnetic,two atoms}.

Since the mode functions of the guided modes are modified by the FBG mirrors, they may contain
narrow resonances. Therefore, we need to perform a special treatment for the contributions from the guided modes. 

We introduce the notation $V_0=V_z$ and $V_{\pm 1}=\mp(V_x\pm i V_y)/\sqrt{2}$ for the spherical components of an arbitrary vector $\mathbf{V}$, and the notation $\mathbf{u}_0=\hat{\mathbf{z}}$ and $\mathbf{u}_{\pm1}=\mp(\hat{\mathbf{x}}\pm i\hat{\mathbf{y}})/\sqrt{2}$ for the spherical basis vectors. 
Without loss of essential physics, we assume that only one spherical component $d_q=d$ of the dipole vector $\mathbf{d}$, where $q=-1$, 0, or 1, is nonzero. 

We use Eq. (\ref{7}) to calculate Eq. (\ref{21a}) and then insert the result into Eq. (\ref{26}). 
We make two approximations. One is to allow the frequency $\omega$ to be negative for the convenience of calculation. The other is that the guided-mode functions $\mathbf{e}^{(\mu)}=\mathbf{e}^{(\omega,f,l)}$ and the factor $\omega \beta'$ are estimated at the atomic transition frequency $\omega_0$.
These approximations are valid because the oscillations described by the exponential factor $e^{-i(\omega-\omega_0)\tau}$ in Eq. (\ref{26}) are generally very fast except for a small region where the mode frequency $\omega$ is close to the atomic transition frequency $\omega_0$. 
As a result, we obtain the following equation for the contribution of the cavity-modified guided modes to the atomic decay:
\begin{eqnarray}\label{28}
\big[\dot{C}_a(t)\big]_{\mathrm{gyd}}&=&
-\frac{\gamma_{\mathrm{gyd}}}{2\pi}\int_{-\infty}^{\infty} G(\omega) d\omega \int_0^t e^{-i(\omega-\omega_0)\tau}
\nonumber\\&&\mbox{}
\times C_a(t-\tau) d\tau.
\end{eqnarray}
Here
\begin{equation}\label{13}
\gamma_{\mathrm{gyd}}=\frac{\omega_0}{2\epsilon_0\hbar v_g}\sum_{fl}
\big|\mathbf{d}\cdot\mathbf{e}^{(\omega_0fl)}\big|^2
=\frac{\omega_0 d^2}{\epsilon_0\hbar v_g}
(|e_{-q}|^2+|e_{q}|^2)
\end{equation}
is the rate of spontaneous emission into guided modes in the absence of the FBG mirrors and
\begin{equation}\label{29}
G(\omega)=
\frac{1+|R|^2+2|R|\cos\Phi\cos(2\beta z)}
{1-|R|^2+4|R|^2 (1-|R|^2)^{-1}\sin^2\Phi} 
\end{equation}
is the cavity impact (enhancement/inhibition) factor. 

In Eq. (\ref{13}), we have introduced the notation $|e_0|=|e_z^{(\omega_0,+,+)}|$ and $|e_{\pm 1}|=(|e_r^{(\omega_0,+,+)}|\mp |e_{\varphi}^{(\omega_0,+,+)}|)/\sqrt{2}$
for the magnitudes of the spherical components of the resonant guided mode functions. We have also introduced the notation $v_g=1/\beta'(\omega_0)$ for the group velocity of the resonant guided field. The cavity-free rate $\gamma_{\mathrm{gyd}}$ of spontaneous emission into guided modes has been calculated and studied in Refs. \cite{cesium decay,magnetic,two atoms}.
 
In Eq. (\ref{29}), we have introduced the notation 
\begin{equation}\label{30}
\Phi=\beta L+\phi_R+(1+q)\pi
\end{equation}
for the shift of the phase of the parallel-to-dipole component of the guided field per cavity crossing with a single reflection. Here $\phi_R$ is the phase of the complex reflection coefficient $R$, that is, $R=|R|e^{i\phi_R}$. Depending on the phase shift per cavity crossing $\Phi$ of the cavity guided field and the axial position $z$ of the atom, the cavity impact factor $G(\omega)$ can be larger or smaller than one, indicating enhancement or inhibition, respectively, of spontaneous emission into guided modes. Such enhancement and inhibition of spontaneous emission are the Purcell effect \cite{Purcell}, which has been studied widely in literature \cite{Berman}. 

Equation (\ref{25}) with the term $\big[\dot{C}_a(t)\big]_{\mathrm{gyd}}$ given by Eq. (\ref{28})
is the basic equation for spontaneous emission of the atom in the model.
We will use this equation to study the emission of the atom in different regimes.

\section{EXPONENTIAL DECAY IN THE OVERDAMPED-CAVITY REGIME}
\label{sec:markov}

We consider the case where the interaction between the atom and the cavity field is weak. 
We assume that the cavity resonance width $\kappa$ is much larger than the characteristic atomic decay rate $\Gamma$, that is, $\kappa\gg\Gamma$. In addition, we assume that the observation time $t$ is much longer than the atomic oscillation period $2\pi/\omega_0$, 
the cavity crossing time $\tau_L=L/v_g$, and the cavity damping time $\kappa^{-1}$, but is much shorter than the atomic decay time $\Gamma^{-1}$, that is, we have $\omega_0t\gg 2\pi$, $t\gg\tau_L$, $\kappa t\gg1$, and $\Gamma t\ll1$. Under these conditions, the Fermi golden rule, which is based on the Born-Markov approximation, is valid \cite{Ching}. We apply the Born-Markov approximation to Eq. (\ref{28}) for the contribution of the cavity-modified guided modes to the atomic decay. In this approximation, we replace $C_a(t-\tau)$ by $C_a(t)$. Furthermore, we extend the upper integration limit $t$ to $+\infty$. Then, we obtain
\begin{equation}\label{57}
\big[\dot{C}_a(t)\big]_{\mathrm{gyd}}=-\frac{\gamma_{\mathrm{cavgyd}}}{2}C_a(t).
\end{equation}
Here
\begin{equation}\label{12}
\gamma_{\mathrm{cavgyd}}=\gamma_{\mathrm{gyd}}G_0
\end{equation}
is the cavity-modified rate of spontaneous emission into guided modes,
with
\begin{equation}\label{15}
G_0=G(\omega_0)=\frac{1+|R|^2+2|R|\cos\Phi_0\cos(2\beta_0 z)}
{1-|R|^2+4|R|^2(1-|R|^2)^{-1} \sin^2\Phi_0} 
\end{equation}
being the resonant cavity impact factor. In Eq. (\ref{15}), we have introduced the notation
$\beta_0=\beta(\omega_0)$ for the propagation constant of the resonant guided light and the notation 
\begin{equation}\label{16}
\Phi_0=\Phi(\omega_0)=\beta_0 L+\phi_R+(1+q)\pi
\end{equation}
for the resonant-light phase shift per cavity crossing with a single reflection.
In deriving Eq. (\ref{57}), we have neglected the contribution of the guided modes to the Lamb shift of the atomic transition frequency.
Note that expression (\ref{15}) for the resonant cavity impact factor $G_0$ is in agreement
with the corresponding results for one-dimensional cavities \cite{Feng,Cook}.

We insert Eq. (\ref{57}) into Eq. (\ref{25}). Then, we obtain the exponential-decay equation
\begin{equation}\label{58}
\dot{C}_a(t)=-\frac{\Gamma}{2}C_a(t),
\end{equation}
with the total atomic decay rate
\begin{equation}\label{59}
\Gamma=\gamma_{\mathrm{cavgyd}}+\gamma_{\mathrm{rad}}=\gamma_{\mathrm{gyd}}G_0+\gamma_{\mathrm{rad}}.
\end{equation}
Thus, in the overdamped-cavity regime, the spontaneous emission of the atom is an exponential decay process. Note that the cavity impact factor $G_0$ and, consequently, the rates $\gamma_{\mathrm{cavgyd}}$ and $\Gamma$ depend on the mirror reflection coefficient $R$. 
They oscillate with varying $z$. They also oscillate with varying cavity length $L$ through their dependences on the phase shift per cavity crossing $\Phi_0$.

The cavity resonance condition is $\Phi_0=m\pi$, where $m$ is an integer number.
Under this resonance condition, Eq. (\ref{15}) for the resonant cavity impact factor $G_0$ reduces to
\begin{equation}\label{17}
G_0=\frac{1+|R|^2+2|R|\cos(2\beta_0 z+m\pi)}{1-|R|^2}. 
\end{equation}
The maximal value
\begin{equation}\label{18}
G_{\mathrm{max}}=\max(G_0)=\frac{1+|R|}{1-|R|} 
\end{equation}
of the factor $G_0$ describes the maximal enhancement of spontaneous emission into guided modes.
The minimal value
\begin{equation}\label{19}
G_{\mathrm{min}}=\min(G_0)=\frac{1-|R|}{1+|R|} 
\end{equation}
of the factor $G_0$ describes the maximal inhibition of spontaneous emission into guided modes.
It is interesting to note that, under the resonance and overdamped-cavity conditions, the maximal enhancement factor $G_{\mathrm{max}}$ and the maximal inhibition factor $G_{\mathrm{min}}$ do not depend on the cavity length $L$. They depend only on the mirror reflection coefficient $R$. 

The above results are different
from the general results for planar Fabry-P\'{e}rot cavities \cite{Martini,Bjork,Dung}.
However, they are in agreement with the results for one-dimensional cavities \cite{Cook,Feng} and 
also with the results for very narrow planar Fabry-P\'{e}rot cavities \cite{Martini}. The reason is that the FBG cavity reflects only the fiber guided modes, which propagate along the fiber axis, and is therefore similar to one-dimensional cavities. It is known that the enhancement factor for a one-dimensional cavity is, in general, larger than that for a corresponding planar Fabry-P\'{e}rot cavity \cite{Bjork}. Therefore, we expect that the FBG cavity can substantially enhance the rate of spontaneous emission into guided modes even when the finesse of the FBG cavity is moderate. Indeed,
for the mirror reflectivity $|R|^2=0.8$ or 0.9, which correspond to the finesse $F=\pi |R|/(1-|R|^2)\cong 14$ or 30, respectively, we obtain the enhancement factor $G_{\mathrm{max}}\cong 18$ or 38, respectively. Such values of the enhancement factor are rather significant.
For comparison, we note that the maximum enhancement factor for a planar Fabry-P\'{e}rot microcavity
with $|R|^2=0.9$ and $L=\lambda_0/2$ is just about 3 \cite{Bjork,Martini}.

%%%%%%%%%%%%%%%%%%%%%%% Figure 2
\begin{figure}[tbh]
\begin{center}
 \includegraphics{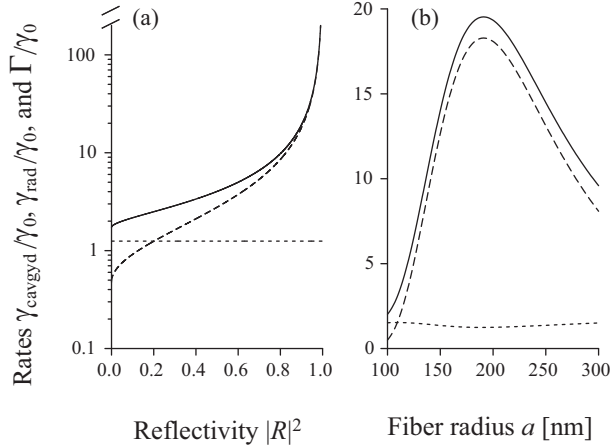}
 \end{center}
\caption{Dependences of the
total spontaneous emission rate $\Gamma$ (solid lines) and its two components 
$\gamma_{\mathrm{cavgyd}}$ (dashed lines) and $\gamma_{\mathrm{rad}}$ (dotted lines) 
on (a) the FBG mirror reflectivity $|R|^2$ and (b) the fiber radius $a$ in the case where 
the dipole of the atom is oriented along one of the spherical basis vectors 
$\mathbf{u}_{\pm1}$. 
In (a), the fiber radius is $a=200$ nm.
In (b), the reflectivity of the FBG mirrors is $|R|^2=0.9$. In both cases,
the atom is located on the fiber surface ($r=a$) and at the cavity center ($z=0$).
The length of the cavity is such as the phase shift per cavity crossing $\Phi_0$ is an even multiple of $\pi$.
The wavelength of the atomic transition is $\lambda_0=852$ nm. 
The refractive indices of the fiber and the vacuum clad are $n_1=1.45$ and $n_2=1$, respectively.
The rates are normalized to the free-space decay rate $\gamma_0$. 
}
\label{fig2}
\end{figure}

The total spontaneous emission rate $\Gamma$ and its components 
$\gamma_{\mathrm{cavgyd}}$ and $\gamma_{\mathrm{rad}}$ depend on the FBG mirror reflectivity $|R|^2$ and the fiber radius $a$. In Fig. \ref{fig2} we plot $\Gamma$, $\gamma_{\mathrm{cavgyd}}$, and $\gamma_{\mathrm{rad}}$ as functions of $|R|^2$ and $a$ in the case where the dipole of the atom is oriented along one of the spherical basis vectors $\mathbf{u}_{\pm1}$. The atom is located on the fiber surface and at the cavity center. The length of the cavity is such that the phase shift per cavity crossing $\Phi_0$ is an even multiple of $\pi$, that is, an even-order resonance is produced. Under this resonance condition, the center of the cavity corresponds to an antinode of the parallel-to-dipole component of the quasistanding-wave guided field formed in the cavity. The rates are normalized to the free-space decay rate $\gamma_0=\omega_0^3d^2/(3\pi\hbar\epsilon_0 c^3)$. Figure \ref{fig2}(a) shows that the cavity-modified rate of spontaneous emission into guided modes $\gamma_{\mathrm{cavgyd}}$ and the total spontaneous emission rate $\Gamma$ increase with increasing reflectivity $|R|^2$. Meanwhile, the rate of spontaneous emission into radiation modes $\gamma_{\mathrm{rad}}$ does not depend on $|R|^2$. In the absence of the cavity ($|R|=0$), the rates of spontaneous emission into guided and radiation modes are $\gamma_{\mathrm{gyd}}=\gamma_{\mathrm{cavgyd}}(|R|=0)\cong0.48\gamma_0$ and $\gamma_{\mathrm{rad}}\cong1.25\gamma_0$, respectively, and the total spontaneous emission rate is $\gamma=\Gamma(|R|=0)\cong1.73\gamma_0$. Figure \ref{fig2}(b) shows that the rate of spontaneous emission into guided modes $\gamma_{\mathrm{cavgyd}}$ and the total spontaneous emission rate $\Gamma$ have a peak when the fiber radius $a$ is around 191 nm.

%%%%%%%%%%%%%%%%%%%%%%% Figure 3
\begin{figure}[tbh]
\begin{center}
 \includegraphics{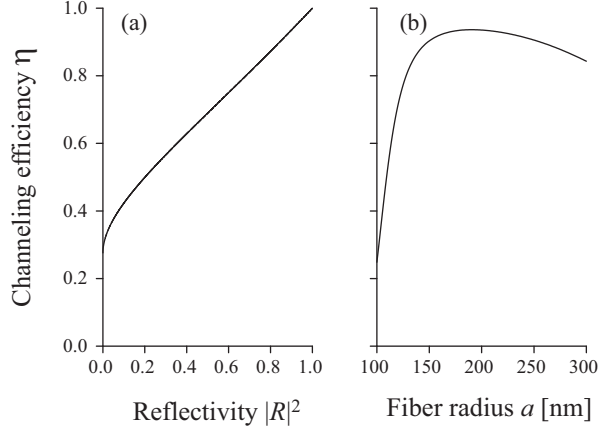}
 \end{center}
\caption{Dependences of the channeling efficiency
 $\eta=\gamma_{\mathrm{cavgyd}}/\Gamma$ 
on (a) the FBG mirror reflectivity $|R|^2$ and (b) the fiber radius $a$ for the parameters of 
Fig. \ref{fig2}.}
\label{fig3}
\end{figure}

The efficiency of channeling of emission into guided modes is characterized by the parameter $\eta=\gamma_{\mathrm{cavgyd}}/\Gamma$. In Fig. \ref{fig3} we plot $\eta$ as functions of the FBG mirror reflectivity $|R|^2$ and the fiber radius $a$ for the parameters of Fig. \ref{fig2}.
Figure \ref{fig3}(a) shows that the channeling efficiency $\eta$ increases with increasing reflectivity $|R|^2$ and can achieve substantial values when $|R|^2$ is close to unity. 
Indeed, for the reflectivity $|R|^2=0.8$ or 0.9, we obtain $\eta\cong 0.87$ (i.e. $87\%$) or $0.94$ (i.e. $94\%$), respectively. Figure \ref{fig3}(b) shows that the channeling efficiency $\eta$ achieves a peak when the fiber radius $a$ is around 191 nm. It is interesting to note that, due to the FBG cavity, the channeling efficiency $\eta$ can achieve substantial values in a relatively wide range of $a$. Indeed, for $|R|^2=0.9$, we find $\eta\geq 80\%$ when $a$ is in the range from 130 to 300 nm.

%%%%%%%%%%%%%%%%%%%%%%% Figure 4
\begin{figure}[tbh]
\begin{center}
 \includegraphics{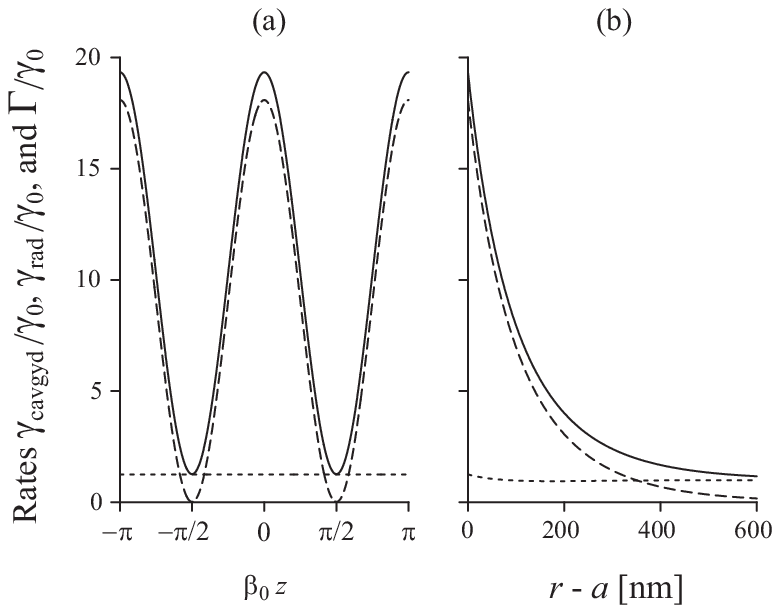}
 \end{center}
\caption{Dependences of the
total spontaneous emission rate $\Gamma$ (solid lines) and its two components 
$\gamma_{\mathrm{cavgyd}}$ (dashed lines) and $\gamma_{\mathrm{rad}}$ (dotted lines) 
on (a) the axial coordinate $z$ and (b) the radial coordinate $r$ of the atom in the case where 
the dipole of the atom is oriented along one of the spherical basis vectors 
$\mathbf{u}_{\pm1}$. 
In (a), the atom is located at $r=a$ (on the fiber surface).
In (b), the atom is located at $z=0$ (at the center of the cavity).
The fiber radius is $a=200$ nm. The reflectivity of the FBG mirrors is $|R|^2=0.9$.
Other parameters are as in Fig. \ref{fig2}.}
\label{fig4}
\end{figure}

The total spontaneous emission rate $\Gamma$ and its components $\gamma_{\mathrm{cavgyd}}$ and $\gamma_{\mathrm{rad}}$ depend on the axial coordinate $z$ and the radial coordinate $r$ of the atom. In Fig. \ref{fig4} we plot $\Gamma$, $\gamma_{\mathrm{cavgyd}}$, and $\gamma_{\mathrm{rad}}$ as functions of $z$ and $r$ in the case where the dipole of the atom is oriented along one of the spherical basis vectors $\mathbf{u}_{\pm1}$. Figure \ref{fig4}(a) shows that the cavity-modified rate of spontaneous emission into guided modes $\gamma_{\mathrm{cavgyd}}$ and the total spontaneous emission rate $\Gamma$ oscillate with varying $z$. The spatial period of the oscillations is $\pi/\beta_0$. The maxima and minima of the rate $\gamma_{\mathrm{cavgyd}}$ or $\Gamma$ correspond to the enhancement and inhibition, respectively, caused by the FBG cavity, and are achieved at the antinodes and nodes, respectively, of the parallel-to-dipole component of the quasistanding-wave guided field formed in the cavity. Meanwhile, the rate of spontaneous emission into radiation modes 
$\gamma_{\mathrm{rad}}$ does not depend on $z$ and is finite. This explains the observation in Fig. \ref{fig4}(a) that, at the nodes of the cavity field, the total atomic decay rate $\Gamma$ remains finite although the component $\gamma_{\mathrm{cavgyd}}$ becomes very small. Figure \ref{fig4}(b) shows that the effect of the fiber on $\Gamma$, $\gamma_{\mathrm{cavgyd}}$, and $\gamma_{\mathrm{rad}}$ is largest for the atom on the fiber surface. It is clear that, when the atom is located at an antinode of the parallel-to-dipole component of the cavity guided field and is near to the fiber surface, $\gamma_{\mathrm{cavgyd}}$ and consequently $\Gamma$ are substantially enhanced by the cavity. When the atom is far away from the fiber ($r\gg a$), the rate $\gamma_{\mathrm{cavgyd}}$ reduces to zero while the rates $\gamma_{\mathrm{rad}}$ and $\Gamma$ approach the free-space value $\gamma_0$.

%%%%%%%%%%%%%%%%%%%%%%% Figure 5
\begin{figure}[tbh]
\begin{center}
 \includegraphics{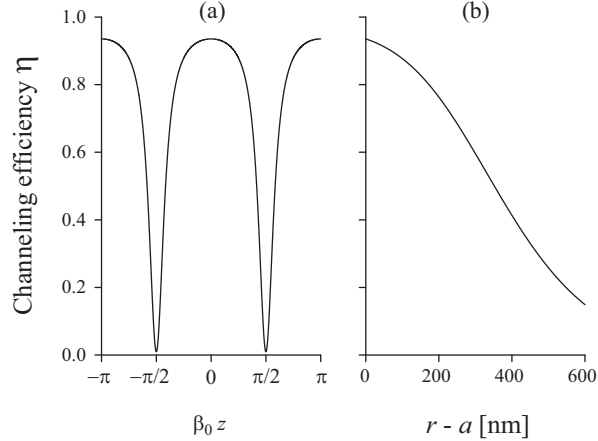}
 \end{center}
\caption{Dependences of the channeling efficiency
 $\eta=\gamma_{\mathrm{cavgyd}}/\Gamma$ 
on (a) the axial coordinate $z$ and (b) the radial coordinate $r$ of the atom for the parameters of Fig. \ref{fig4}.}
\label{fig5}
\end{figure}

We plot in Fig. \ref{fig5} the efficiency of channeling of
emission into guided modes $\eta=\gamma_{\mathrm{cavgyd}}/\Gamma$ 
against the axial coordinate $z$ and the radial coordinate $r$ of the atom 
for the case of Fig. \ref{fig4}. Figure \ref{fig5}(a) shows that $\eta$ oscillates with varying $z$, with the period $\pi/\beta_0$. It is interesting to note that $\eta$ is substantial in broad 
regions around the antinodes of the parallel-to-dipole component of the cavity guided field
and has narrow dips at the nodes. The appearance of such features is due to the fact that the total atomic decay rate $\Gamma$ has two components, one is enhanced or inhibited around the antinodes or nodes, respectively, and the other is not modified by the cavity and is substantial. 
Figure \ref{fig5}(b) shows that the channeling efficiency $\eta$ reduces with increasing atom-to-surface distance $r-a$ and is substantial in a broad region of $r-a$. Indeed, more than 50\% of emission can be directed into guided modes when the atom-to-surface distance is less than 350 nm.
In addition, the channeling efficiency $\eta$ can be significant even when $r-a$ is large. 
Indeed, up to about 15\% of emission can be directed into guided modes when the atom-to-surface distance is 600 nm.

%%%%%%%%%%%%%%%%%%%%%%% Figure 6
\begin{figure}[tbh]
\begin{center}
 \includegraphics{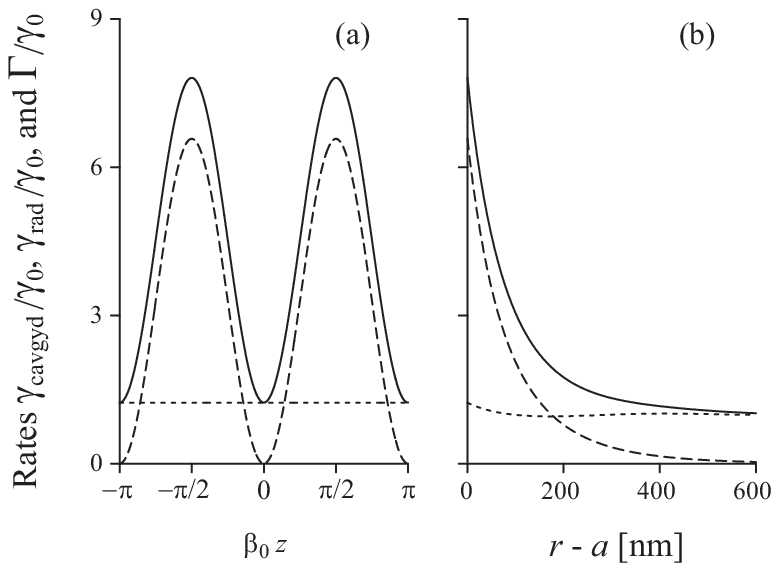}
 \end{center}
\caption{Dependences of the
total spontaneous emission rate $\Gamma$ (solid lines) and its two components 
$\gamma_{\mathrm{cavgyd}}$ (dashed lines) and $\gamma_{\mathrm{rad}}$ (dotted lines) 
on (a) the axial coordinate $z$ and (b) the radial coordinate $r$ of the atom in the case where 
the dipole of the atom is oriented along the fiber axis $z$. 
In (a), the atom is located at $r=a$ (on the fiber surface).
In (b), the atom is located at $\beta_0z=\pm\pi/2$ (one-fourth of the guided-light wavelength from the cavity center).
The fiber radius is $a=200$ nm. The reflectivity of the FBG mirrors is $|R|^2=0.9$.
Other parameters are as in Fig. \ref{fig2}.}
\label{fig6}
\end{figure}

%%%%%%%%%%%%%%%%%%%%%%% Figure 7
\begin{figure}[tbh]
\begin{center}
 \includegraphics{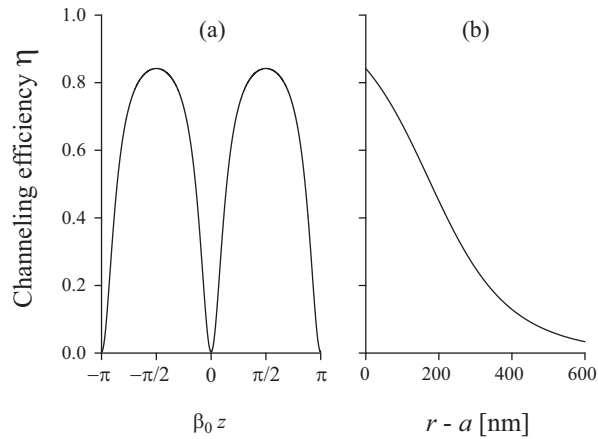}
 \end{center}
\caption{Dependences of the channeling efficiency
 $\eta=\gamma_{\mathrm{cavgyd}}/\Gamma$ 
on (a) the axial coordinate $z$ and (b) the radial coordinate $r$ of the atom for the parameters of Fig. \ref{fig6}.}
\label{fig7}
\end{figure}

The numerical results presented in Figs. \ref{fig2}--\ref{fig5} were obtained for the case where the dipole vector $\mathbf{d}$ of the atom is perpendicular to the fiber axis $z$. 
Meanwhile, the spontaneous decay rates and their modifications caused by the FBG cavity 
depend on the orientation of the atomic dipole. 
We plot in Figs. \ref{fig6} and \ref{fig7} the spatial dependences of the decay rates $\Gamma$, $\gamma_{\mathrm{cavgyd}}$, and $\gamma_{\mathrm{rad}}$
and the channeling efficiency $\eta$ for the case where the dipole of the atom is oriented along the fiber axis $z$. Comparison between Figs. \ref{fig4} and \ref{fig6}
and between Figs. \ref{fig5} and \ref{fig7} shows that the decay rates and the channeling efficiency
are smaller for an atom with a dipole parallel to the fiber axis than for an atom with a dipole perpendicular to the fiber axis. In addition, we observe that the positions of maxima (minima) in the case of Figs. \ref{fig4}(a) and \ref{fig5}(a) correspond to the positions of minima (maxima) in the case of Figs. \ref{fig6}(a) and \ref{fig7}(a). Such opposite behaviors are due to the differences
between the phase shifts per reflection of the longitudinal ($q=0$) and the transverse ($q=\pm1$) components of the guided field.

\section{DELAY-DIFFERENTIAL EQUATION FOR MULTIPLE REFLECTIONS}
\label{sec:delay}

We now examine Eq. (\ref{25}) in the case where the coupling between the atom and the cavity-modified guided modes may be strong and, consequently, the Born-Markov approximation for the contribution of the cavity guided modes to the atomic decay may not be valid.
We follow the approach of Refs. \cite{Cook,Feng,Dung} and derive a delay-differential equation that describes explicitly multiple reflections in our model.

We start from Eq. (\ref{29}) for the cavity impact factor $G(\omega)$.
We expand the denominator of the fraction in this equation into a Fourier series as
\begin{eqnarray}\label{31}
\lefteqn{\frac{1}{1-|R|^2+4|R|^2 (1-|R|^2)^{-1}\sin^2\Phi}} 
\nonumber\\&& 
=\frac{2}{1+|R|^2}\sum_{n=0}^{\infty}\frac{|R|^{2n}}{1+\delta_{n,0}}\cos(2n\Phi).
\end{eqnarray}
With the help of the above formula, we expand the cavity impact factor $G(\omega)$ into a series as
\begin{eqnarray}\label{32}
\lefteqn{G(\omega)=2\sum_{n=0}^{\infty}\frac{|R|^{2n}}{1+\delta_{n,0}}\cos(2n\Phi)
+\sum_{n=0}^{\infty}|R|^{2n+1}}
\nonumber\\&&\mbox{}
\times\big\{\cos[(2n+1)\Phi+2\beta z]+\cos[(2n+1)\Phi-2\beta z]\big\}. 
\nonumber\\
\end{eqnarray}
We insert Eq. (\ref{32}) into Eq. \eqref{28} and calculate the integrals with the help of the formulas
\begin{eqnarray}\label{34}
\lefteqn{\int_{-\infty}^{\infty}\cos(2n\Phi) e^{-i(\omega-\omega_0)\tau} d\omega}
\nonumber\\&&\qquad
=(1+\delta_{n,0})\pi e^{2ni\Phi_0}\delta(\tau-2n\tau_L),
\nonumber\\
\lefteqn{\int_{-\infty}^{\infty}\cos[(2n+1)\Phi \pm 2\beta z] e^{-i(\omega-\omega_0)\tau} d\omega}
\nonumber\\&&\qquad 
=\pi e^{(2n+1)i\Phi_0}e^{\pm 2i\beta_0 z}\delta(\tau-2n\tau_L-\tau_{\pm}),
\end{eqnarray}
where $\tau_{+}=(L+2z)/v_g$ and $\tau_{-}=(L-2z)/v_g$ are the position-dependent group delays due to the left and right mirrors, respectively, and $\tau_L=L/v_g=(\tau_{+}+\tau_{-})/2$ is the group delay per cavity crossing. In deriving expressions (\ref{34}) we have neglected the group velocity dispersion. When insert the result of the calculations into Eq. (\ref{25}), we obtain
\begin{eqnarray}\label{35}
\lefteqn{\dot{C}_a(t)
= -\frac{\gamma_{\mathrm{gyd}}}{2}\bigg\{
C_a(t)\Theta(t)+2\sum_{n=1}^{\infty}|R|^{2n}e^{2ni\Phi_0}}
\nonumber\\&&\mbox{}
\times C_a(t-2n\tau_L)\Theta(t-2n\tau_L)
+\sum_{n=0}^{\infty}|R|^{2n+1}e^{(2n+1)i\Phi_0}
\nonumber\\&&\mbox{}
\times \Big[e^{2i\beta_0 z}
C_a(t-2n\tau_L-\tau_{+}) \Theta(t-2n\tau_L-\tau_{+})
\nonumber\\&&\mbox{}
+e^{-2i\beta_0 z}C_a(t-2n\tau_L-\tau_{-}) \Theta(t-2n\tau_L-\tau_{-})\Big]\bigg\}
\nonumber\\&&\mbox{}
-\frac{\gamma_{\mathrm{rad}}}{2}C_a(t).
\end{eqnarray}
Here $\Theta(t)$ is the Heaviside step function, whose value is zero for negative argument and one for positive argument.

Equation (\ref{35}) is a delay-differential equation for the
decay of an atom near a fiber with a pair of FBG mirrors. The first term, $C_a(t)\Theta(t)$,
does not depend on the reflection coefficient $R$. This term
describes spontaneous emission into guided modes in the absence of the cavity. 
The other terms are associated with the coefficients of the type $R^n$, where $n=1,2,\dots$. 
Such terms describe the backaction of the emitted photon on the atom after the photon is reflected from the mirrors $n$ times. The quantities of the type $2n\tau_L$ and $2n\tau_L+\tau_{\pm}$ 
are the group delays. The factors of the type $e^{2ni\Phi_0}$ and
$e^{(2n+1)i\Phi_0}e^{\pm 2i\beta_0z}$ describe the phase shifts of the parallel-to-dipole component of the guided field due to the propagation along the nanofiber and the reflection from the FBG gratings. Thus, the delay-differential equation (\ref{35}) describes spontaneous emission of the atom in terms of multiple reflections. Due to the effect of retardation on the atomic state, the atomic decay may become nonexponential. We note that the absorption of the guided field by the fiber material can be incorporated into the theory by adding an imaginary part to the longitudinal wave number $\beta_0$, which appears in expression (\ref{16}) for the phase shift per cavity crossing $\Phi_0$ and also in the local phase factors $e^{\pm 2i\beta_0z}$. 

It is clear from Eq. (\ref{35}) that, when $t<\tau_{\mathrm{min}}\equiv\min \{\tau_L,\tau_{+},\tau_{-}$\}, we have $\dot{C}_a=-(\gamma/2) C_a$. Here, $\gamma=\gamma_{\mathrm{gyd}}+\gamma_{\mathrm{rad}}$
is the total rate of spontaneous emission into both types of modes in the absence of the cavity.
The above result means that the atom does not feel the presence of the cavity until the time $t=\tau_{\mathrm{min}}$. 

In the framework of the Born-Markov approximation, we can replace the variables $C_a(t-2n\tau_L)$ and $C_a(t-2n\tau_L-\tau_{\pm})$ in Eq. (\ref{35}) by $C_a(t)$. Then, Eq. (\ref{35}) reduces to $\dot{C}_a=-(\Gamma/2) C_a$. Here, $\Gamma=\gamma_{\mathrm{gyd}}G_0+\gamma_{\mathrm{rad}}$ is
the total rate of spontaneous emission into both types of modes in the presence of the cavity.
This result is in agreement with the results of Sec. \ref{sec:markov} on the exponential decay of the atom in the overdamped-cavity regime [see Eqs. (\ref{58}) and (\ref{59})].

The delay-differential equation \eqref{35} is similar to but different from the corresponding
equation for the case of planar Fabry-P\'{e}rot microcavities \cite{Dung}. The key difference is that the coefficients in the delay-differential equation for planar Fabry-P\'{e}rot microcavities contain $\xi^{-1}$-,
$\xi^{-2}$-, and $\xi^{-3}$-terms which correspond to the dipole radiation field, the induced field, and
the electrostatic field, respectively, due to the mirror images \cite{Dung}. Here, $\xi=2n\omega_0\tau_L$, $2n\omega_0\tau_L+\omega_0\tau_+$, or $2n\omega_0\tau_L+\omega_0\tau_-$ is the retardation time. 
The absence of the $\xi^{-1}$-, $\xi^{-2}$-, and $\xi^{-3}$-terms in the expressions for the coefficients in Eq. (\ref{35}) is because the FBG cavity reflects only the guided modes 
and is therefore similar to one-dimensional cavities. Due to this reason, Eq. (\ref{35}) is almost
the same as the corresponding equation for one-dimensional cavities \cite{Cook,Feng}.
A difference between the two cases is that Eq. (\ref{35}) contains an additional term, namely
the term $-(\gamma_{\mathrm{rad}}/2)C_a(t)$, which describes spontaneous emission from the atom into radiation modes. Another difference is that the cross-section area of the cavity modes is rigorously included in the expression for the rate $\gamma_{\mathrm{gyd}}$ of spontaneous emission into guided modes but is phenomenologically included in the treatment of Ref. \cite{Cook} or is omitted in the treatment of Ref. \cite{Feng}.

%%%%%%%%%%%%%%%%%%%%%%% Figure 8
\begin{figure}[tbh]
\begin{center}
 \includegraphics{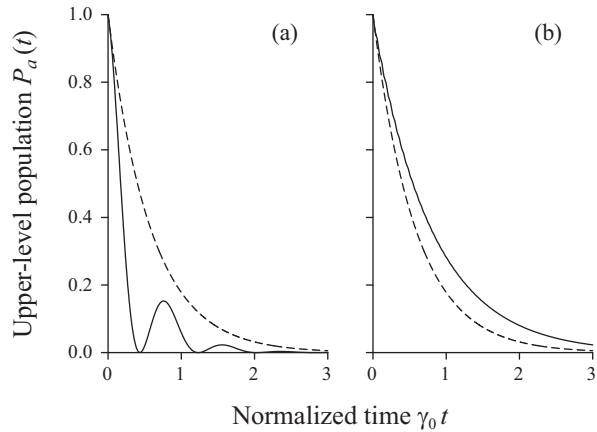}
 \end{center}
\caption{Time evolution of the upper-state population $P_a=|C_a|^2$ of the atom in a long FBG cavity.
The length of the cavity is $L=20$ cm and is tuned to resonance with the atomic
transition frequency so that the phase shift per cavity crossing $\Phi_0$ is an (a) even or (b) odd multiple of $\pi$. The reflectivity of the FBG mirrors is $|R|^2=0.9$.
The atom is located on the fiber surface and at the cavity center.
Other parameters are as in Fig. \ref{fig2}.
For comparison, the exponential decay of the atomic upper-state 
population in the absence of the cavity is shown by the dashed curves.
}
\label{fig8}
\end{figure}

The delay-differential equation \eqref{35} for the probability amplitude $C_a$ of the atomic upper state $|a\rangle$ can be solved numerically \cite{Dung,Droplet} by using a subroutine developed in Ref. \cite{Hairer}.
We solve this equation and plot in Fig. \ref{fig8} the time evolution of the atomic upper-state population $P_a=|C_a|^2$ for the case where the FBG cavity length is $L= 20$ cm.
The atom is located on the fiber surface ($r=a$) and at the cavity center ($z=0$).
The cavity length is tuned to resonance with the atomic transition frequency so that
the phase shift per cavity crossing $\Phi_0$ is an even or odd multiple of $\pi$, that is,
the center of the cavity corresponds to an antinode or a node, respectively, of the parallel-to-dipole component of the cavity guided field. Other parameters are as in Fig. \ref{fig2}. For comparison, the exponential decay of the atomic upper-state population in the absence of the cavity is shown by the dashed curves. The solid line in Fig. \ref{fig8}(a) shows the occurrence of vacuum Rabi oscillations \cite{Cook,Feng,Dung}. Such oscillations are due to strong coupling between the atom and the guided field in the FBG cavity. It is interesting to note that strong coupling
and vacuum Rabi oscillations can occur even when the cavity length is large ($L=20$ cm)
and the finesse of the cavity is moderate [$F=\pi |R|/(1-|R|^2)\cong30$]. There are two reasons for this. The first reason is that the field in the guided modes of the nanofiber is confined in a small area of the transverse plane, that is, the guided-mode cross-section area is small. Due to this reason,
the effective cavity-mode volume can be small and, consequently, the cavity--atom coupling constant can be large even when the FBG cavity length is large. The other reason is that the FBG cavity is similar to one-dimensional cavities. In such a cavity, the cavity damping rate reduces with increasing cavity length faster than the strength of the coupling between the atom and the cavity. 
Unlike one-dimensional cavities, planar Fabry-P\'{e}rot optical cavities have off-axis modes, which reduce the cavity QED effects \cite{Bjork,Dung}. In addition, the radius of the cavity mode in a planar Fabry-P\'{e}rot optical cavity increases with increasing cavity length $L$ and with increasing
mirror reflectivity $|R|^2$ \cite{Bjork}. In the case of curved Fabry-P\'{e}rot optical cavities, the typical values of the mode waist are much larger the wavelength of light. Consequently, the realization of strong coupling in a planar or a curved Fabry-P\'{e}rot cavity requires a smaller cavity length and a higher finesse than in a nanofiber-based cavity. The typical Fabry-P\'{e}rot optical cavities used in experimental realizations of strong coupling have lengths
in the range from 10 to 100 $\mu$m and finesse factors on the order of $10^5$
\cite{Kimble group,Rempe,Shimizu,McKeever,Maunz,Sauer}. 

The solid line in Fig. \ref{fig8}(b) shows the decay of the atom is almost exponential. Comparison between the solid and the dashed lines shows that the atomic decay is slightly slowed down by presence of the cavity. In the case of this figure, the atom is positioned at a node of the cavity guided field
and, therefore, spontaneous emission of the atom into guided modes is substantially inhibited. 
The total spontaneous emission of the atom is mainly determined by spontaneous emission into radiation modes. This decay channel is slightly weaker than the cavity-free atomic decay.

%%%%%%%%%%%%%%%%%%%%%%% Figure 9
\begin{figure}[tbh]
\begin{center}
 \includegraphics{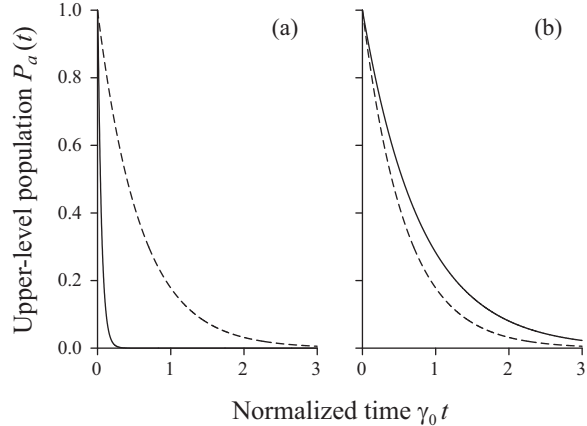}
 \end{center}
\caption{Time evolution of the upper-state population $P_a=|C_a|^2$ of the atom in a short FBG cavity.
The length of the cavity is $L=2$ mm and is tuned to resonance with the atomic
transition frequency so that the phase shift per cavity crossing $\Phi_0$ is an (a) even or (b) odd multiple of $\pi$. The reflectivity of the FBG mirrors is $|R|^2=0.9$.
The atom is located on the fiber surface and at the cavity center.
Other parameters are as in Figs. \ref{fig2} and \ref{fig8}.
For comparison, the exponential decay of the atomic upper-state 
population in the absence of the cavity is shown by the dashed curves.
}
\label{fig9}
\end{figure}

We plot in Fig. \ref{fig9} the time evolution of the atomic upper-state population $P_a=|C_a|^2$ for the case where the FBG cavity length is $L= 2$ mm. Other parameters are as in Figs. \ref{fig2}
and \ref{fig8}. The figure shows that the atomic population decay is almost exponential. 
According to Fig. \ref{fig9}(a), the exponential decay of the atom at an antinode of the cavity guided field (see the solid curve) is substantially faster than the cavity-free atomic decay (see the dashed curve). According to Fig. \ref{fig9}(b), the exponential decay of the atom at a node of the cavity guided field (see the solid curve) is slightly slower than the cavity-free atomic decay (see the dashed curve). 

Comparison between Figs. \ref{fig8}(a) and \ref{fig9}(a) shows that vacuum Rabi oscillations can be observed in the case of Fig. \ref{fig8}(a), where
the cavity length is rather large ($L=20$ cm), but not in the case of Fig. \ref{fig9}(a), where the cavity length is much shorter ($L=2$ mm). Thus, vacuum Rabi oscillations cannot occur when the cavity length is too short. This result is different from the common belief that the smaller cavity can produce the stronger vacuum Rabi oscillations \cite{Berman}. Such a belief was based on the results for high-finesse microcavities.
Meanwhile, our model involves the use of a moderate-finesse nanofiber-based cavity.

Comparison between Figs. \ref{fig8}(b) and \ref{fig9}(b) shows that the time dependences of the atomic upper-state population $P_a$ in the two cases are essentially the same. Moreover, they are almost identical to the exponential decay of the atom into radiation modes. The reason is the following: in the two cases, the atom is positioned at a node of the cavity guided field and, hence, spontaneous emission into guided modes is inhibited. Since this effect is substantial enough, the total atomic decay process is mainly determined by the process of spontaneous emission into radiation modes.

%%%%%%%%%%%%%%%%%%%%%%% Figure 10
\begin{figure}[tbh]
\begin{center}
 \includegraphics{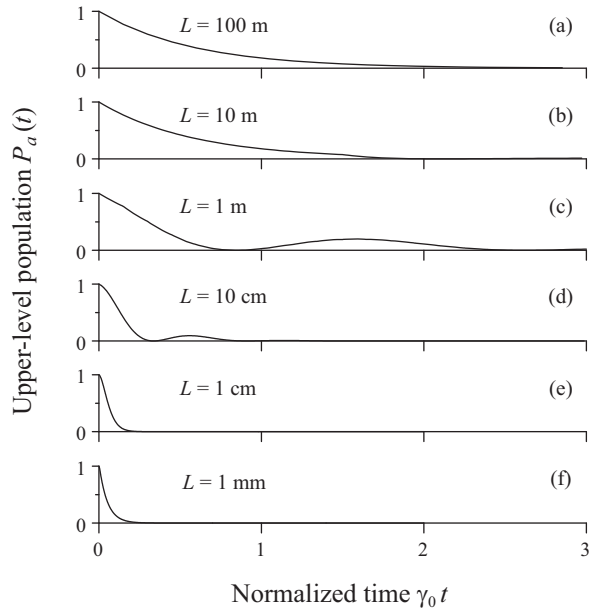}
 \end{center}
\caption{Time evolution of the upper-state population $P_a=|C_a|^2$ of the atom in a FBG cavity
with length $L=100$ m (a), 10 m (b), 1 m (c), 10 cm (d), 1 cm (e), and 1 mm (f).
The length of the cavity is tuned to exact resonance with the atomic
transition frequency so that the phase shift per cavity crossing $\Phi_0$ is an even multiple of $\pi$.
The atom is located on the fiber surface and at the cavity center.
The reflectivity of the FBG mirrors is $|R|^2=0.9$. 
Other parameters are as in Fig. \ref{fig2}.
}
\label{fig10}
\end{figure}

We plot in Fig. \ref{fig10} the time evolution of the atomic upper-state population $P_a=|C_a|^2$ for different values of the cavity length, in the range from 100 m to 1 mm. The length of the cavity is tuned to exact resonance with the atomic transition frequency so that the phase shift per cavity crossing $\Phi_0$ is an even multiple of $\pi$, that is, the center of the cavity corresponds to an antinode of the parallel-to-dipole component of the cavity guided field. Since the cavity length is rather large in the cases of Figs. \ref{fig10}(a)--(c), we take into account the absorption of the guided light by the fiber material in the calculations. For this purpose, we add an imaginary part of $\alpha/2$, with $\alpha=10^{-5}$ cm$^{-1}$ being the typical absorption coefficient for silica, to the longitudinal wave number $\beta_0$, which appears in Eq. (\ref{35}) through the phase shifts $\Phi_0$ and $\pm 2\beta_0z$. However, we neglect the nonradiative atomic decay caused by the material absorption  \cite{Yeung,Knol}. The ratio of the rate $\gamma_{\mathrm{nonrad}}$ of such a nonradiative process to the natural decay rate $\gamma_0$ is given in the limit of small atom-to-surface distances $r-a$ by the factor $\epsilon_I/[2|\epsilon+1|^2k_0^3(r-a)^3]$, where $\epsilon_I$ is the imaginary part of the complex permitivity $\epsilon$ \cite{Yeung,Knol}. In the case of silica, $\epsilon_I$ is on the order of $10^{-10}$. Therefore, the nonradiative decay rate $\gamma_{\mathrm{nonrad}}$ of an atom with the transition wavelength $\lambda_0=852$ nm of the cesium $D_2$ line is significant only when the distance $r-a$ from the atom to the fiber surface is on the order of or less than 0.2 \AA. 
Such a threshold distance is very small as compared to the light wavelength and the fiber radius, and is even smaller than the Bohr radius. Therefore, it is neglected in our treatment. 
The aim of the choice of the value $r-a=0$ for the calculations of Fig. \ref{fig10} as well as Figs. \ref{fig8} and \ref{fig9} is to show the most dramatic effects in the limiting case where the effects of the material absorption, the surface-induced potential, and the surface roughness on the atomic decay can be neglected. Our additional calculations, not shown here, confirm that the numerical results presented in Figs. \ref{fig8}--\ref{fig10} remain basically unchanged when the value zero for $r-a$ is replaced by a few nanometers.

Figure \ref{fig10}(a) shows that, when the cavity length $L$ is large enough, the decay of the atomic upper-state population $P_a$ is almost exponential. Such a decay is close to the exponential decay of the atom in the absence of the cavity. The cavity-free atomic decay rate is $\gamma=\gamma_{\mathrm{gyd}}+\gamma_{\mathrm{rad}}$ and is approximately equal to $1.73\gamma_0$ in the case considered. We note that, in the case of large $L$, we may also observe vacuum Rabi oscillations. However, such oscillations are weak, and the corresponding period is large and approaches the cavity crossing time $\tau_L$. 

Figures \ref{fig10}(c) and \ref{fig10}(d) show that significant vacuum Rabi oscillations occur when the cavity length $L$ is in a range on the order of 10 cm to 1 m. Such lengths are rather large. We mention again that, in the case of planar \cite{Martini,Bjork,Dung} and curved
\cite{Berman,Thompson,Mabuchi,Kimble group,Rempe,Shimizu,McKeever,Maunz,Sauer} Fabry-P\'{e}rot cavities, due to the substantial magnitudes of the cavity-mode cross-section areas and the effects of the off-axis modes, strong coupling cannot be realized in long cavities. 
Comparison between Fig. \ref{fig10}(c) and \ref{fig10}(d) shows that a decrease in the cavity length leads to a decrease in the vacuum Rabi period. This feature is in agreement with the fact that the vacuum Rabi frequency $\Omega$ is proportional to the factor $1/\sqrt{\tau_L}=\sqrt{v_g/L}$ [see Eq. (\ref{44})], which characterizes the cavity mode density or the inverse of the cavity mode volume.

Figure \ref{fig10}(f) shows that, when the cavity length $L$ is small enough, the decay of the atomic upper-state population $P_a$ returns to the exponential-decay regime, with a cavity-modified decay rate $\Gamma=\gamma_{\mathrm{gyd}}G_0+\gamma_{\mathrm{rad}}$. We find $\Gamma\cong 19.33\gamma_0\cong
11.20 \gamma$ in the case of the figure. 

%%%%%%%%%%%%%%%%%%%%%%% Figure 11
\begin{figure}[tbh]
\begin{center}
 \includegraphics{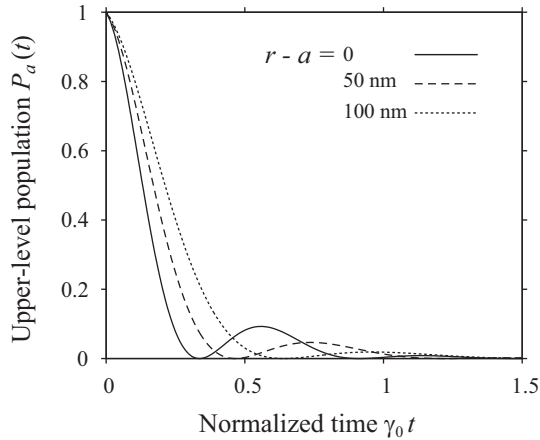}
 \end{center}
\caption{Time evolution of the upper-state population $P_a=|C_a|^2$ of the atom
at different distances $r-a=0$ (solid line), 50 nm (dashed line), and 100 nm (dotted line) from the fiber surface in a FBG cavity. The cavity length is $L=10$ cm. The atom is located at the center of the cavity. Other parameters are as in Figs. \ref{fig2} and \ref{fig10}.
}
\label{fig11}
\end{figure}

The dependence of the time evolution of the atomic upper-state population $P_a=|C_a|^2$ 
on the distance $r-a$ from the atom to the fiber surface is illustrated in Fig. \ref{fig11}.
The figure shows clearly that vacuum Rabi oscillations can be observed even when the distance $r-a$ is as large as 100 nm. The strong coupling between such a distant atom and the guided field is due to the effect
of the FBG cavity.

We conclude this section by presenting an analytical solution to the delay-differential equation (\ref{35}) in a particular case where the atom is at the center of the cavity, i.e., $z=0$. In this case, Eq. (\ref{35}) reduces to
\begin{eqnarray}\label{35a}
\dot{C}_a(t)&=& -\gamma_{\mathrm{gyd}}\sum_{n=1}^{\infty}|R|^{n}e^{ni\Phi_0}
C_a(t-n\tau_L)\Theta(t-n\tau_L)
\nonumber\\&&\mbox{}
-\frac{\gamma}{2}C_a(t).
\end{eqnarray}
The above equation has been solved analytically in Ref. \cite{Dung}. 
When we follow the result of Ref. \cite{Dung}, we find
\begin{eqnarray}
C_a(t)&=& e^{-\gamma t/2} \sum_{n=0}^{\infty}
|R|^n e^{ni\Phi_0} e^{n\gamma\tau_L/2} \Theta(t-n\tau_L)
\nonumber\\&&\mbox{}
\times\sum_{k_1,\dots,k_n}
(-\gamma_{\mathrm{gyd}})^p\frac{(t-n\tau_L)^p} 
{k_1!k_2!\cdots k_n!},
\label{sol}
\end{eqnarray}
where the inner sum is over all non-negative integers $k_1, k_2,
\cdots , k_n$ that satisfy the condition $k_1+2k_2+\cdots+nk_n=n$, and $p=k_1+k_2+\cdots+k_n$ is their sum. Expression (\ref{sol}) describes the time dependence of
the probability amplitude $C_a$ of the atomic upper state $|a\rangle$. 
The results of calculations of expression (\ref{sol}) are in complete agreement with the numerical solutions of Eq. (\ref{35a}).

\section{SINGLE-MODE CAVITY}
\label{sec:single mode}

In order to get insight into our model, we approximate the delay-differential equation \eqref{35} under the single-mode cavity condition. For this purpose, we follow the procedures of Refs. \cite{Cook,Feng,Dung}. We consider a cavity mode, whose frequency $\omega_c$ is determined by the resonance condition $\Phi(\omega_c)=m\pi$. Here $m$ is an integer number. We introduce the parameter
\begin{equation}\label{39}
\Delta=\frac{\Phi(\omega_c)-\Phi(\omega_0)}{\tau_L}=\frac{m\pi-\Phi_0}{\tau_L}\cong \omega_c-\omega_0,
\end{equation}
which characterizes the detuning of the cavity mode frequency $\omega_c$ from the atomic transition frequency $\omega_0$. It is clear that the separation between the cavity-mode frequencies $\omega_c$ is $\Delta\omega_c\cong\pi/\tau_L$. We rewrite the delay-differential equation \eqref{35} as 
\begin{eqnarray}\label{38}
\lefteqn{\dot{C}_a(t)
=-\frac{\gamma_{\mathrm{gyd}}}{2}\bigg\{
2\sum_{n=1}^{\infty} e^{-2n(i\Delta+\kappa /2)\tau_L}C_a(t-2n\tau_L)}
\nonumber\\&&\mbox{}
\times \Theta(t-2n\tau_L)
+(-1)^m\sum_{n=0}^{\infty}e^{-(2n+1)(i\Delta+\kappa /2)\tau_L}
\nonumber\\&&\mbox{}
\times \Big[e^{2i\beta_0 z}
C_a(t-2n\tau_L-\tau_{+}) \Theta(t-2n\tau_L-\tau_{+})
\nonumber\\&&\mbox{}
+e^{-2i\beta_0 z}C_a(t-2n\tau_L-\tau_{-}) \Theta(t-2n\tau_L-\tau_{-})\Big]\bigg\}
\nonumber\\&&\mbox{}
-\frac{\gamma}{2}C_a(t),
\end{eqnarray}
where 
\begin{equation}\label{41}
\kappa=\frac{2}{\tau_L}\big|\ln|R|\big|
\end{equation}
is the cavity damping rate. Since the frequency separation between the cavity modes is $\Delta\omega_c\cong\pi/\tau_L$, the cavity finesse is approximately given by $F=\Delta\omega_c/\kappa\cong \pi/(2|\ln |R||)$. 
 
Although Eq. \eqref{38} is valid for an arbitrary integer number $m$, we choose such an integer number $m$ for which the mode frequency $\omega_c$ is closest to the atomic transition frequency $\omega_0$, that is, the detuning $\Delta$ is smallest. We consider the case where the condition $|\Delta|\tau_L\ll1$ is satisfied. This condition means that $|\Delta|\ll \Delta\omega_c$, that is, the cavity-atom detuning $\Delta$ is much smaller than the cavity-mode frequency separation $\Delta\omega_c$. In this case, the effect of the cavity mode with the frequency $\omega_c$ on spontaneous emission of the atom is dominant over that of other cavity modes. 
Furthermore, we assume that $|R|\cong1$, so we have $\kappa \tau_L\ll1$. 
In addition, we assume that $t\gg \tau_L$ and $\gamma\tau_L\ll1$. 
Under the above conditions, we can replace the sums in Eq. \eqref{38}
by integrals and hence obtain 
\begin{eqnarray}\label{42}
\lefteqn{\dot{C}_a(t)=-\frac{\gamma_{\mathrm{gyd}}}{\tau_L}\cos^2(\beta_c z+m\pi/2)}
\nonumber\\&&\mbox{}
\times\int\limits_{0}^{t} e^{-(i\Delta+\kappa /2)(t-t')}C_a(t')dt'
-\frac{\gamma}{2}C_a(t),
\end{eqnarray}
where $\beta_c=\beta(\omega_c)\cong\beta_0+\Delta/v_g$.
When we differentiate the above equation with respect to $t$, 
we find the second-order differential equation
\begin{equation}\label{43}
\ddot{C}_a+\Big(i\Delta+\frac{\kappa +\gamma}{2}\Big)\dot{C}_a
+\bigg[\frac{\Omega^2}{4}+\Big(i\Delta+\frac{\kappa }{2}\Big)\frac{\gamma}{2}\bigg]C_a=0,
\end{equation}
where
\begin{equation}\label{44}
\Omega=2\sqrt{\frac{\gamma_{\mathrm{gyd}}}{\tau_L}
}|\cos(\beta_c z+m\pi/2)|
\end{equation}
is the cavity--atom coupling constant. Note that Eqs. (\ref{43}) and (\ref{44}) are in agreement with the results for high-finesse one-dimensional cavities \cite{Berman,Cook,Feng}.
It is clear from Eq. (\ref{44}) that the cavity--atom coupling constant $\Omega$ is inversely proportional to the factor $\sqrt{\tau_L}=\sqrt{L/v_g}$, which effectively characterizes the cavity mode volume or the inverse of the cavity mode density. In addition, $\Omega$ is proportional to the rate of spontaneous emission into guided modes $\sqrt{\gamma_{\mathrm{gyd}}}$. Since the field in guided modes is tightly confined in the transverse plane, that is, the guided-mode cross-section area is small, the rate $\gamma_{\mathrm{gyd}}$ can be substantial when the atom is close to the fiber surface \cite{cesium decay}. Therefore, $\Omega$ can achieve substantial values even when the cavity length $L$ is large. 

We analyze the case of exact cavity--atom resonance, where $\Delta=0$, that is, $\omega_c=\omega_0$ and, consequently, $\beta_c=\beta_0$. In this case, Eq. \eqref{43} reduces to
\begin{equation}\label{45}
\ddot{C}_a+\frac{\kappa +\gamma}{2}\dot{C}_a+\frac{\Omega^2+\kappa \gamma}{4}C_a=0.
\end{equation}
The initial conditions for the spontaneous emission process are $C_a(0)=1$ and $\dot{C}_a(0)=-\gamma/2$. For these initial conditions, the solution to Eq. \eqref{45} is found to be
\begin{equation}\label{47}
C_a(t)=e^{-(\kappa +\gamma)t/4 }\Big[\cosh(\Lambda t/2)+\frac{\kappa -\gamma}{2\Lambda }\sinh(\Lambda t/2)\Big],
\end{equation}
where
\begin{equation}\label{48}
\Lambda =\sqrt{(\kappa -\gamma)^2/4-\Omega^2}.
\end{equation}
Below, we study several different regimes of the general solution (\ref{47}).

First, we consider the strong-coupling (underdamped-cavity) regime, where 
$\Omega$ is sufficiently large that the condition 
\begin{equation}\label{63}
2\Omega\gg\kappa,\gamma 
\end{equation}
is satisfied. In this regime, Eq. (\ref{47}) yields
\begin{equation}\label{49}
C_a(t)\cong e^{-(\kappa +\gamma)t/4 }\cos(\Omega t/2).
\end{equation}
Hence, the population $P_a(t)= |C_a(t)|^2$ of the atomic upper state $|a\rangle$ is found to be
\begin{equation}\label{50}
P_a(t)\cong e^{-(\kappa +\gamma)t/2 }\cos^2(\Omega t/2).
\end{equation}
The above solution describes the occurrence of vacuum Rabi oscillations in the strong-coupling regime \cite{Cook,Feng,Dung}. 

The strong-coupling condition (\ref{63}) can be rewritten as
\begin{equation}\label{61}
L_2\gg L\gg L_1,
\end{equation}
where
\begin{eqnarray}\label{62}
L_2&=&\frac{16v_g\gamma_{\mathrm{gyd}}}{\gamma^2} \cos^2(\beta_0 z+m\pi/2),
\nonumber\\
L_1&=&\frac{v_g}{4\gamma_{\mathrm{gyd}}}\frac{\ln^2 |R|}{\cos^2(\beta_0 z+m\pi/2)}.
\end{eqnarray}
Condition (\ref{61}) says that the strong-coupling regime can be realized only if the cavity length 
$L$ is sufficiently small as compared to $L_2$ and is sufficiently large as compared to $L_1$.
It is clear that condition (\ref{61}) can be realized only if 
\begin{equation}\label{61a}
L_2\gg L_1. 
\end{equation}

When the atom is positioned at a node of the parallel-to-dipole component of the cavity guided field, we have $\cos(\beta_0 z+m\pi/2)=0$, which leads to $L_2=0$, $L_1=\infty$, and $\Omega=0$. In this case, condition (\ref{61a}) and the strong-coupling condition (\ref{63}) cannot be satisfied.

When the atom is positioned at an antinode, we have $\cos(\beta_0 z+m\pi/2)=\pm1$, which leads to 
\begin{equation}\label{67}
\Omega=2\sqrt{\frac{\gamma_{\mathrm{gyd}}}{\tau_L}}
\end{equation}
and
\begin{eqnarray}\label{64}
L_2&=&\frac{16v_g\gamma_{\mathrm{gyd}}}{\gamma^2},
\nonumber\\
L_1&=&\frac{v_g}{4\gamma_{\mathrm{gyd}}}\ln^2 |R|\cong \frac{\pi^2 v_g}{16F^2\gamma_{\mathrm{gyd}}}.
\end{eqnarray}
In this case, condition (\ref{61a}) can be rewritten as 
\begin{equation}\label{65}
\frac{\gamma_{\mathrm{gyd}}}{\gamma}\gg \frac{\big|\ln|R|\big|}{8}
\cong\frac{\pi}{16F}.
\end{equation}
Condition (\ref{65}) can be satisfied if the cavity-free channeling efficiency factor 
$\gamma_{\mathrm{gyd}}/\gamma=\gamma_{\mathrm{gyd}}/(\gamma_{\mathrm{gyd}}+\gamma_{\mathrm{rad}})$ 
and the cavity finesse $F$ are sufficiently substantial. Under condition (\ref{65}), we can choose an appropriate cavity length $L$ that satisfies condition (\ref{61a}) for the strong-coupling regime at an antinode.

Since the field in guided modes of the nanofiber is confined in a small area of the transverse plane, that is, the guided-mode cross-section area is small, the channeling efficiency factor $\gamma_{\mathrm{gyd}}/\gamma$ can achieve substantial values when the atom is close to the fiber surface \cite{cesium decay}. In this case, condition (\ref{65}) can be satisfied for moderate values of the finesse $F$ of the cavity. Furthermore, the cavity--atom coupling constant $\Omega$ and the upper limit value $L_2$ can be large. Consequently, the strong-coupling condition (\ref{63}) and its equivalent form (\ref{61}) can be satisfied for large values of the cavity length $L$. It is interesting to note that, in our model, the cavity damping rate $\kappa$, given by Eq. (\ref{41}), decreases faster with increasing $L$ than the cavity--atom coupling constant $\Omega$, given by Eq. (\ref{67}). 
Due to this fact, the upper limit $L_2$ for condition (\ref{61}) is determined by the requirement $2\Omega\gg\gamma$ but not by the requirement $2\Omega\gg\kappa$. This is a common feature of one-dimensional cavities \cite{Cook,Feng}. We emphasize again that, in the case of planar \cite{Martini,Bjork,Dung} and curved \cite{Berman,Thompson,Mabuchi,Kimble group,Rempe,Shimizu,McKeever,Maunz,Sauer} Fabry-P\'{e}rot cavities, due to the substantial magnitudes of the cavity mode cross-section areas and the effects of the off-axis modes, strong coupling cannot be realized in large cavities. The typical lengths of Fabry-P\'{e}rot optical cavities used in experimental realizations of strong coupling are in the range from 10 to 100 $\mu$m \cite{Kimble group,Rempe,Shimizu,McKeever,Maunz,Sauer}. Such short cavities
must have high finesse in order to achieve the strong-coupling regime.

Unlike the upper limit value $L_2$ for the strong-coupling condition (\ref{61}), the lower limit value $L_1$ is determined by the requirement $2\Omega\gg\kappa$. In the case of high-finesse cavities, where $F\gg1$, we have $L_1\to 0$. However, when $F$ is moderate, $L_1$ can become large. Thus, strong coupling cannot be realized in a FBG cavity with a moderate finesse $F$ if the cavity length $L$ is too short. In such a cavity, the cavity damping rate $\kappa$ is much larger than the cavity--atom coupling constant $\Omega$. 

We discuss the possibilities of strong coupling and consequential vacuum Rabi oscillations in the cases
of Figs. \ref{fig8}(a) and \ref{fig10}. In the cases of these figures, the atom is positioned at an antinode of the cavity guided field, and the mirror reflectivity is $|R|^2=0.9$ (the cavity finesse is
$F\cong30$). For the parameters of these figures, we find the critical values $L_2\cong 17$ m and $L_1\cong 1$ cm. It is clear that condition (\ref{61}) is satisfied in the case of Fig. \ref{fig8}(a), where $L=20$ cm, and in the cases of Figs. \ref{fig10}(c) and \ref{fig10}(d), where the cavity length is $L=1$ m and 10 cm, respectively. This explains why vacuum Rabi oscillations are observed in the above-mentioned figures. Furthermore, we obtain $(\Omega,\kappa,\gamma)/\gamma_0\cong(7.97,3.51,1.73)$ in the case of Fig. \ref{fig8}(a), $(\Omega,\kappa,\gamma)/\gamma_0\cong(3.56,0.70,1.73)$ in the case of Fig. \ref{fig10}(c), and $(\Omega,\kappa,\gamma)/\gamma_0\cong(11.27,7.02,1.73)$ in the case of Fig. \ref{fig10}(d). These parameters satisfy the strong-coupling condition (\ref{63}). For the free-space atomic decay rate $\gamma_0=5.2$ MHz of the cesium $D_2$ line, the cavity--atom coupling constant achieves the values $\Omega\cong$ 42, 19, and 59 MHz in the cases of Figs. \ref{fig8}(a), \ref{fig10}(c), and \ref{fig10}(d), respectively. Such values of $\Omega$ are comparable to the values obtained
in the experiments on realization of strong coupling in high-finesse Fabry-P\'{e}rot optical microcavities \cite{Kimble group,Rempe,Shimizu,McKeever,Maunz,Sauer}.

Next, we consider the overdamped-cavity regime, where the condition
\begin{equation}\label{68} 
\kappa \gg 2 \Omega,\gamma
\end{equation} 
is satisfied. In this regime, Eq. (\ref{47}) yields
\begin{equation}\label{51}
C_a(t)\cong e^{-\Gamma t/2}
\end{equation}
and, hence, we obtain
\begin{equation}\label{52}
P_a(t)= |C_a(t)|^2\cong e^{-\Gamma t}.
\end{equation}
Here
\begin{equation}\label{52a}
\Gamma=\gamma_{\mathrm{gyd}}G_0+\gamma_{\mathrm{rad}}
\end{equation}
is the total decay rate of the atom, with
\begin{equation}\label{53}
G_0=1+\frac{\Omega^2}{\kappa\gamma_{\mathrm{gyd}}}=
1+\frac{2}{\big|\ln|R|\big|}\cos^2(\beta_0 z+m\pi/2)
\end{equation}
being the cavity impact factor for the rate of spontaneous emission into guided modes.
The maximal enhancement factor is
\begin{equation}\label{54}
G_{\mathrm{max}}=1+\frac{2}{\big|\ln|R|\big|}\cong 1+\frac{4F}{\pi}
\end{equation}
and the maximal inhibition factor is
\begin{equation}\label{55}
G_{\mathrm{min}}=1.
\end{equation}
We note that Eqs. (\ref{53})--(\ref{55}) agree with Eqs. (\ref{17})--(\ref{19}) of 
Sec. \ref{sec:markov} in the limit $|R|\to1$.

The overdamped-cavity condition (\ref{68}) can be rewritten as
\begin{equation}\label{69} 
L\ll L_1,L_3,
\end{equation}
where
\begin{equation}\label{70} 
L_3=\frac{2v_g}{\gamma}\big|\ln|R|\big| \cong \frac{\pi v_g}{F\gamma}.
\end{equation} 
Condition (\ref{69}) indicates that the overdamped-cavity regime, where the spontaneous emission of the atom is an exponential-decay process with a cavity-modified rate $\Gamma$, can be realized only when the FBG cavity is sufficiently short. When the finesse $F$ of the cavity is moderate, $L_1$ and $L_3$ can be large. For the parameters of Figs. \ref{fig9} and \ref{fig10}, we find $L_1\cong 1$ cm and $L_3\cong41$ cm. Then, the overdamped-cavity condition (\ref{68}) becomes $L\ll 1$ cm. It is clear that the case of Fig. \ref{fig9}, where $L=2$ mm, and the case of Fig. \ref{fig10}(f), where $L=1$ mm, correspond to the overdamped-cavity regime.

Finally, we discuss the case where $L\gg L_2,L_3$. In this case, we have $\gamma\gg\Omega,\kappa$. Then, Eq. (\ref{47}) yields $C_a(t)=e^{-\gamma t/2}$ and, hence, we find $P_a(t)=e^{-\gamma t}$. Thus, when the cavity is very long, the upper-state population $P_a$ of the atom reduces exponentially with the cavity-free atomic decay rate $\gamma$. Such a decay is observed in Fig. \ref{fig10}(a) although the parameters for this figure do not satisfy the conditions $t\gg \tau_L$ and $\gamma\tau_L\ll1$, which were used in deriving Eq. (\ref{43}) from Eq. (\ref{38}).

\section{Summary}
\label{sec:summary}

We have studied spontaneous emission of an atom near a nanofiber with two fiber Bragg grating (FBG) mirrors. We have shown that the coupling between the atom and the guided modes of the nanofiber can be significantly enhanced by the FBG cavity even when the cavity finesse is moderate. We have found that, when the fiber radius is 200 nm and the cavity finesse is about 30, up to 94\% of spontaneous emission from the atom can be channeled into the guided modes in the overdamped-cavity regime. 

We have derived a delay-differential equation which explicitly describes the effects of multiple reflections of the guided field on the atom. We have analyzed this equation in different regimes of the atomic decay. We have shown numerically and analytically that vacuum Rabi oscillations and strong coupling can occur in the FBG cavity even when the cavity finesse is moderate (about 30) and the cavity length is large (on the order of 10 cm to 1 m), unlike the case of planar and curved Fabry-P\'{e}rot cavities. We have identified two reasons for this possibility. One reason is that the field in the guided modes of the nanofiber is confined in a small area of the transverse plane. Due to this reason, the effective cavity-mode volume can be small even when the FBG cavity length is large. Another reason is that the FBG cavity is similar to one-dimensional cavities, where there are no off-axis modes.

\appendix

\section{Mode functions of the fundamental guided modes of a nanofiber}
\label{sec:guided}

For the fundamental guided modes, the propagation constant $\beta$ is determined by the
fiber eigenvalue equation \cite{fiber books}
\begin{eqnarray}
\frac{J_0(h a)}{h a J_1(h a)}&=&
-\frac{n_1^2+n_2^2}{2n_1^2}\frac{K_1'(q a)}{q a K_1(q a)}+ \frac{1}{h^2 a^2}
\nonumber\\&&\mbox{}
-\Bigg[\left(\frac{n_1^2-n_2^2}{2n_1^2}\frac{K_1'(q a)}{q a K_1(q a)}\right)^2
\nonumber\\&&\mbox{}
+\frac{\beta^2}{n_1^2 k^2}\left(\frac{1}{q^2a^2}+\frac{1}{h^2a^2}\right)^2 \Bigg]^{1/2}.
\label{1}
\end{eqnarray}
Here the parameters $h=(n_1^2k^2-\beta^2)^{1/2}$ and $q=(\beta^2-n_2^2k^2)^{1/2}$ characterize the fields inside and outside the fiber, respectively. The notation $J_n$ and $K_n$ stand for the Bessel functions of the first kind and the modified Bessel functions of the second kind, respectively. 

The mode functions of the electric parts of the fundamental guided modes \cite{fiber books} are given, 
for $r<a$, by
\begin{eqnarray}
e_{r}^{(\mu)}&=&iC\frac{q}{h}\frac{K_1(qa)}{J_1(ha)}[(1-s)J_0(hr)-(1+s)J_2(hr) ],
\nonumber\\
e_{\varphi}^{(\mu)}&=&-lC\frac{q}{h}\frac{K_1(qa)}{J_1(ha)}[(1-s)J_0(hr)+(1+s)J_2(hr) ],
\nonumber\\
e_{z}^{(\mu)}&=& fC\frac{2q}{\beta}\frac{K_1(qa)}{J_1(ha)}J_1(hr),
\label{2}
\end{eqnarray}
and, for $r>a$, by
\begin{eqnarray}
e_{r}^{(\mu)}&=&iC[(1-s)K_0(qr)+(1+s)K_2(qr) ],
\nonumber\\
e_{\varphi}^{(\mu)}&=&-lC[(1-s)K_0(qr)-(1+s)K_2(qr) ],
\nonumber\\
e_{z}^{(\mu)}&=& fC\frac{2q}{\beta}K_1(qr).
\label{3}
\end{eqnarray} 
Here the parameter $s$ is defined as 
$s=({1}/{q^2a^2}+{1}/{h^2a^2})/[{J_1^\prime (ha)}/{haJ_1(ha)}+{K_1^\prime (qa)}/{qaK_1(qa)}]$, and 
the coefficient $C$ is determined from the normalization condition 
\begin{equation}
\int _{0}^{2\pi}d\varphi\int _{0}^{\infty}n_{\mathrm{rf}}^2\,|\mathbf{e}^{(\mu)}|^2r\,dr=1.
\label{4}
\end{equation}
Here $n_{\mathrm{rf}}(r)=n_1$ for $r<a$, and $n_{\mathrm{rf}}(r)=n_2$ for $r>a$.

\section{Mode functions of the radiation modes of a nanofiber}
\label{sec:radiation}

For the radiation modes, we have $-kn_2<\beta<kn_2$.
The characteristic parameters for the field in the inside and outside of the fiber are $h=\sqrt{k^2n_1^2-\beta^2}$ and $q=\sqrt{k^2n_2^2-\beta^2}$, respectively.
The mode functions of the electric parts of the radiation modes $\nu=(\omega\beta m l)$
\cite{fiber books} are given, for $r<a$, by
\begin{eqnarray}\label{A37}
e_r^{(\nu)}&=&
\frac{i}{h^2}\left[\beta hAJ'_m(hr)+im\frac{\omega\mu_0}{r}BJ_m(hr)\right],\nonumber\\ 
e_{\varphi}^{(\nu)}&=&
\frac{i}{h^2}\left[im\frac{\beta}{r}AJ_m(hr)-h\omega\mu_0BJ'_m(hr)\right],\nonumber\\
e_z^{(\nu)}&=&AJ_m(hr),
\end{eqnarray}
and, for $r>a$, by 
\begin{eqnarray}\label{A38}
e_r^{(\nu)}&=&
\frac{i}{q^2}\sum_{j=1,2}
\left[\beta q C_jH^{(j)\prime}_m(qr)+im\frac{\omega\mu_0}{r}D_jH^{(j)}_m(qr)\right],\nonumber\\
e_{\varphi}^{(\nu)}&=&
\frac{i}{q^2}\sum_{j=1,2}
\left[im\frac{\beta}{r}C_jH^{(j)}_m(qr)-q\omega\mu_0D_jH^{(j)\prime}_m(qr)\right], \nonumber\\
e_z^{(\nu)}&=&\sum_{j=1,2}C_jH_m^{(j)}(qr).
\end{eqnarray}
The coefficients $C_j$ and $D_j$ are related to the coefficients $A$ and $B$ as 
\cite{Tromborg}
\begin{eqnarray}\label{A39}
C_j&=&(-1)^{j}\frac{i\pi q^2a}{4n_2^2}(AL_j+i\mu_0cBV_j),\nonumber\\
D_j&=&(-1)^{j-1}\frac{i\pi q^2a}{4}(i\epsilon_0cAV_j-BM_j),
\end{eqnarray}
where
\begin{eqnarray}\label{A40}
V_j&=&\frac{mk\beta}{ah^2q^2}
(n_2^2-n_1^2)
J_m(ha)H_m^{(j)*}(qa),\nonumber\\
M_j&=&\frac{1}{h}J'_m(ha)H_m^{(j)*}(qa)
-\frac{1}{q}J_m(ha)H_m^{(j)*\prime}(qa),\nonumber\\
L_j&=&\frac{n_1^2}{h}J'_m(ha)H_m^{(j)*}(qa)
-\frac{n_2^2}{q}J_m(ha)H_m^{(j)*\prime}(qa).\nonumber\\
\end{eqnarray}
We specify two polarizations by choosing $B=i\eta A$ and $B=-i\eta A$ for $l=+$
and $l=-$, respectively.
The orthogonality of the modes requires
\begin{eqnarray}\label{A41}
&&\int _0^{2\pi}d\varphi\int _{0}^{\infty}n_{\mathrm{rf}}^2
\left[\mathbf{e}^{(\nu)}\mathbf{e}^{(\nu')*}\right]_{\beta=\beta',m=m'}
\;rdr \nonumber\\&&
=N_{\nu}\delta_{ll'}\delta(\omega-\omega').
\end{eqnarray}
This leads to
\begin{equation}\label{A42}
\eta=\epsilon_0c\sqrt{\frac{n_2^2|V_j|^2+|L_j|^2}{|V_j|^2+n_2^2|M_j|^2}}.
\end{equation}
The normalization constant $N_{\nu}$ is given by 
\begin{equation}\label{A43}
N_{\nu}=\frac{8\pi \omega}{q^2}\left(n_2^2|C_j|^2+\frac{\mu_0}{\epsilon_0}|D_j|^2\right).
\end{equation}

\end{document}